\documentclass[12pt]{article}
\usepackage{a4,a4wide}

\usepackage{amscd}\usepackage{amsmath,latexsym,amsmath,amssymb,amstext}
\usepackage{amssymb} 
\usepackage{amsfonts}

\usepackage[debug,colorlinks=true
,urlcolor=blue
,anchorcolor=blue
,citecolor=green
,filecolor=blue
,linkcolor=blue
,menucolor=blue
,pagecolor=blue
,linktocpage=true,pageanchor=false
]{hyperref}

\usepackage{epsfig}
\usepackage{graphicx}

\usepackage{enumerate}
\usepackage{fancybox}

\def\Tr{\mathrm{Tr}}

\def\half{{1\over2}}
\def\nn{\nonumber\\}

\def\sgn{\text{sgn}}

\def\[{\left[}
\def\]{\right]}
\def\({\left(}
\def\){\right)}

\newcommand{\wt}{\widetilde}

\def\={\stackrel{\bullet}{=}}
\def\nn{\notag\\}

\def\half{{1\over2}}

\def\[{\left[}
\def\]{\right]}
\def\({\left(}
\def\){\right)}

\def\cC{{\cal C}}
\def\cD{{\cal D}}

\def\cO{{\cal O}}

\def\cS{{\cal S}}

\def \be {\begin{equation}}
\def \ee {\end{equation}}
\def \bea {\begin{eqnarray}}
\def \eea {\end{eqnarray}}
\def \beal#1 {\begin{align}#1\end{align}}
\def\bes#1{\begin{equation}\begin{split}#1\end{split}\end{equation}}
\def \nn {\notag\\}

\def\aver#1{\left\langle #1 \right\rangle}

\begin{document}

\setlength{\topmargin}{-.5in} 
\setlength{\textheight}{9.in}
\baselineskip = 17pt

\makeatletter
\renewcommand{\theequation}{%
\thesection.\arabic{equation}}
\@addtoreset{equation}{section}
\makeatother

\begin{titlepage}
\vspace{-3cm}
\title{
\begin{flushright}
\normalsize{ 
TIFR/TH/15-19\\
July 2015}
\end{flushright}
       \vspace{1.5cm}
       Chern Simons Bosonization along RG Flows
       \vspace{1.5cm}
}

\author{
\!\!\!\!\!\!\!\! 
Shiraz Minwalla$^{a),1}$, 
Shuichi Yokoyama$^{b),2}$
\\[25pt] 
\!\!\!\!\!\!\!\!{\it \normalsize $^{a)}$Department of Theoretical Physics, Tata Institute of Fundamental Research,}\\
\!\!\!\!\!\!\!\!{\it \normalsize Homi Bhabha Road, Mumbai 400005, India}\\
\!\!\!\!\!\!\!\!{\it \normalsize $^{b)}$Physics Department, Technion - Israel Institute of Technology,}\\
\!\!\!\!\!\!\!\!{\it \normalsize Technion City -
Haifa 3200003}
\\[10pt]
\!\!\!\!\!\!\!\!{\small \tt E-mail: $^1$minwalla(at)theory.tifr.res.in, ${}^2$shuichitoissho(at)gmail.com}
}

\date{}
\maketitle

\thispagestyle{empty}

\vspace{.2cm}

\begin{abstract}
\vspace{0.3cm}
\normalsize
It has previously been conjectured that the theory of free fundamental scalars 
minimally coupled to a Chern Simons gauge field is dual to the theory of 
critical fundamental fermions minimally coupled to a level rank dual Chern 
Simons gauge field.
In this paper we study RG flows away from these two fixed points by turning
on relevant operators. In the 't Hooft large $N$ limit we compute the 
thermal partition along each of these 
flows and find a map of parameters under which the two partition functions 
agree exactly with each other all the way from the UV to the IR. We 
conjecture that the bosonic and fermionic RG flows are dual to each other under this map of parameters. Our flows can 
be tuned to end at the gauged critical scalar theory and gauged 
free fermionic theories respectively. Assuming the validity of our conjecture, 
this tuned trajectory may be viewed as RG flow from the gauged theory of free bosons to the gauged theory of free fermions.

\end{abstract}
\end{titlepage}


 
\section{Introduction}

$U(N_B)$ invariant relativistic quantum field theories with $N_B$ complex 
scalar fields have two well known fixed points in three dimensions. 
The first is the theory of $N_B$ free complex massless scalars. The second is the $U(N_B)$ 
invariant $N_B$ component generalization of the Wilson-Fisher fixed point \cite{Wilson:1971dc}, the 
so called theory of critical scalars. In a similar manner, the space of 
$U(N_F)$ invariant quantum field theories with $N_F$ complex fermions also 
plausibly admits two known conformal fixed points. The first is a system of $N_F$ massless free
fermions. At least in the $\frac{1}{N_F}$ expansion, there is also a second 
fixed point which we will refer to as the theory of critical fermions \cite{Rosenstein:1988pt}. (See also \cite{Arefeva:1980ms}.) 

The free scalar theory has two relevant operators, while the critical 
scalar theory has a single relevant operator. It is well known that there 
exists an RG flow from the theory of free scalars to the theory 
of critical scalars. 
In a similar manner the 
critical fermion theory has two relevant operators while the free
fermion theory has one. At least in the large $N$ limit there exists and RG flow 
from the critical to the free fermion theory.

The free and critical conformal field theories 
described above each admit one discrete parameter generalization that 
enriches their dynamics  \cite{Giombi:2011kc,Aharony:2011jz}. 
This generalization is obtained by gauging the
$U(N_B) / U(N_F)$ global symmetry groups with a level $\kappa_B/\kappa_F$
Chern Simons coupled gauge field.
In the rest of this paper we will refer to these theories as 
the regular and critical scalar/fermion theories respectively.
At least in the 't Hooft large $N$ limit 
all four resultant theories continue to be conformal
\cite{Giombi:2011kc,Aharony:2011jz}. Moreover the structure 
of RG flows between these fixed points is expected to be qualitatively 
unaffected by gauging (of course the integers $\kappa_B$ and $\kappa_F$, like 
$N_B$ and $N_F$, are invariant under RG flows).\footnote{
Let  $k_B$ and $k_F$ denote the levels of the boundary dual WZW theories. 
We define 
$\kappa_{B/F}=k_{B/F} + {\rm sgn} (k_{B/F} ) N_{B/F}$. We also define 
$\lambda_{B/F}=\frac{N_{B/F}}{\kappa_{B/F}}$, which is fixed under the large $N_{B/F}$ limit. Notice that $|N_{B/F}|\leq|\kappa_{B/F}|$ and thus 
$|\lambda_{B/F}|\leq 1$.  }

In the 't Hooft large $N$ limit these four gauged theories (the free and
critical bosonic theory together with the free and critical fermionic 
theory) are interacting three dimensional conformal field theories that nonetheless appear to be exactly solvable at all values of the 
't Hooft coupling $\lambda$. 
We now have several exact results for correlation functions of local gauge 
invariant operators \cite{Giombi:2011rz,Maldacena:2011jn,Maldacena:2012sf,Aharony:2012nh,GurAri:2012is, Frishman:2013dvg,  Bardeen:2014qua, Gurucharan:2014cva,
Frishman:2014cma, Moshe:2014bja,
 Bedhotiya:2015uga}, thermal partition functions  
\cite{Giombi:2011kc, Jain:2012qi, Yokoyama:2012fa, Aharony:2012ns,  Jain:2013py,
Jain:2013gza, Takimi:2013zca, Yokoyama:2013pxa} (see also 
\cite{Banerjee:2012gh,Banerjee:2013nca}) and S matrices 
\cite{Jain:2014nza,Dandekar:2014era,Inbasekar:2015tsa}
in these and related theories. 

One of the most interesting patterns to have emerged from the exact solutions 
of these theories is the observation that the regular/ critical bosonic 
theory appear to be 
dual to the critical/regular fermionic theory. The existence of 
a three dimensional {\it bosonization} duality of this nature was first 
suggested in \cite{Giombi:2011kc} motivated partly by the conjectured bulk 
Vasiliev duals of these theories. Substantial direct field theory evidence 
for such dualities was obtained in 
\cite{Maldacena:2012sf}; the first concrete proposal 
for such a duality including a proposed map between parameters 
between the dual pairs was presented in \cite{Aharony:2012nh}. 
The original duality conjectures were generalized to a broader range 
of theories in \cite{Aharony:2012ns, 
Jain:2013gza} and a great deal of additional evidence for these dualities 
was obtained in several of the papers cited in the previous paragraph. 

The duality map between the regular/critical bosonic and critical/regular 
fermionic theories has an extremely simple structure. In the large
$N$ limit the regular/critical bosonic theory at rank $N_B$ and level 
$\kappa_B$ is conjectured to be dual to the critical/regular fermionic theory 
at rank $N_F$ and level $\kappa_B$ with $\kappa_F=-\kappa_B$, 
$N_F=|\kappa_B|-N_B$. In terms of the 't Hooft coupling, the duality map takes
the form $\lambda_F=\lambda_B-{\rm sgn}(\lambda_B)$.

If two conformal field theories are exactly dual to each other, then the 
RG flows away from these field theories must also be dual to each other. 
The conjectured duality between the regular bosonic and critical fermionic 
theories thus makes an immediate prediction; the two parameter set of RG flows 
away from the gauged regular bosonic theory must be dual to the two parameter 
set of RG flows away from the gauged critical fermionic theory. In 
particular the RG flow from the gauged regular boson to the gauged critical 
boson theory must be dual to the RG flow from the gauged critical fermion 
to the gauged regular fermion theory. In other words the known duality 
between the end points of these two RG flows should lift to a duality 
between the two flows as function of scale, all the way from the UV to the IR. 
In this paper we will find direct evidence for this claim.  

In the large $N$ limit, the discussion of the previous paragraph
 may be generalized as follows. 
In addition to the two marginal deformations, regular scalar and critical 
fermion theories each possesses an operator that is irrelevant at any 
finite $N$ but is exactly marginal in the strict large $N$ limit 
(its $\beta$ function turns out to be of order $\frac{1}{N}$ \cite{Aharony:2011jz}). In the strict large $N$ limit, therefore, the conjectured duality of 
the regular scalar and critical fermion theories implies a duality between 
the the three parameter family of theories obtained by deforming these 
theories with arbitrary proportions of the two relevant and one marginal 
operators. As the marginal operator is a strict large $N$ artifact it is a
little artificial to turn it on. Nonetheless in the rest of this paper we 
will work with the full three parameter set of theories on both sides of 
the duality, simply because we can. Readers interested only in RG flows that 
have a finite $N$ counterpart are advised to restrict attention 
to the the appropriate two parameter set of theories by setting 
$x_6=1$ and $y_6=0$ in all formulae below.\footnote{Moreover this 
marginal operator turns out to be irrelevant about the IR fixed point 
of the RG flows and so drops out of the IR dynamics of 
these flows.}

In this paper we find direct evidence for the duality between the 
three parameter set of quantum field theories obtained from relevant 
and marginal deformations of the regular scalar and critical fermionic 
theories in the 't Hooft large  $N$ limit. We proceed as follows.  
We follow \cite{Giombi:2011kc, Aharony:2012ns, Jain:2012qi,Jain:2013py,
Jain:2013gza} to use large $N$ techniques to compute the 
thermal partition function of the scalar theory defined by the Euclidean 
Lagrangian%
\footnote{We have implicitly assumed the contraction of gauge indices, 
for example $\bar\phi \phi = \bar\phi_m \phi^m$ where $m$ is the fundamental gauge index.} 
\beal{
S_B  &= \int d^3 x  \biggl[i \varepsilon^{\mu\nu\rho}{\kappa_B\over 4\pi}
\Tr( A_\mu\partial_\nu A_\rho -{2 i\over3}  A_\mu A_\nu A_\rho)
+ D_\mu \bar \phi D^\mu\phi  \nn
&~~~~~~~~~~~~~
+m_B^2 \bar\phi \phi +  {4\pi b_4 \over \kappa_B}(\bar\phi \phi)^2
+ {(2\pi)^2x_6 \over (\kappa_B)^2} (\bar\phi \phi)^3\biggl]
\label{rst}
}
where the gauge covariant derivative acts the scalar fields as 
\be
D_{\mu} \phi =(\partial _{\mu} -i A_\mu ) \phi, \quad 
D_{\mu} \bar \phi =(\partial _{\mu} \bar \phi + i  \bar \phi A_\mu ).
\label{covariantderivative}
\ee
We also independently compute the thermal partition of the fermionic theory 
defined by the Lagrangian  
\beal{~\label{csfnonlinear}
 S_F =&\int d^3 x \bigg[ i \varepsilon^{\mu\nu\rho} {\kappa_F \over 4 \pi}
\Tr( A_\mu\partial_\nu A_\rho -{2 i\over3}  A_\mu A_\nu A_\rho)
+  \bar{\psi} \gamma_\mu D^{\mu} \psi \nn
&~~~~~~~~~~~~~
+\sigma_F (\bar\psi \psi - {\kappa_Fy_2^2\over4\pi} ) - {\kappa_F y_4\over4\pi} \sigma_F^2 +  {\kappa_F y_6 \over 4\pi} \sigma_F^3\bigg]. 
}
We then demonstrate that the thermal partition functions of these 
two theories agree under the identifications
\be
\lambda_F=\lambda_B-{\rm sgn}(\lambda_B), \quad
y_6 ={1-x_6 \over 4}, \quad  
y_4 = b_4, \quad y_2^2 = m_B^2
\label{dualitytransform}
\ee
(see the next section for a discussion of how this is done.)%
\footnote{ When the chemical potential for $U(1)$ flavor current are introduced for each theories, they interchange under the duality map \cite{Aharony:2012ns,Jain:2013gza}. 
Although the generalization to include the chemical potential is straight-forward, we do not consider them in this paper for simplicity. 
}
Our results lead us to conjecture that the theories \eqref{rst} and 
\eqref{csfnonlinear} are dual to each other in the strict large $N$ limit, 
and suggest a similar duality at large but finite $N$ (upon restricting  
attention to $x_6=1$ and $y_6=1$).\footnote{
Note that \eqref{dualitytransform} proposes a linear map between the 
Lagrangians of the two theories, in contrast with the highly nonlinear 
map obtained in the study of a more complicated system in  \cite{Jain:2013gza}.
We do not understand the underlying reason behind the extreme simplicity of the 
transformation rules \eqref{dualitytransform}.
 }

We will now explain the interpretation of these results in terms of RG flows.
Let us first note that the duality map \eqref{dualitytransform} maps the 
bosonic theory with $x_6=1$, $b_4=m_B^2=0$ to the fermionic theory with 
$y_6=y_4=y_2^2=0$. This is simply a restatement of the duality between 
the gauged regular bosonic theory and the gauged critical fermionic theory 
discussed in detail above. 

The parameters $x_6-1$, $b_4$ and $m_B^2$ parametrize deformations away from
the regular scalar theory while $y_6, y_4$ and $y_2^2$ parametrize deformations 
away from 
the critical fermion theory. At linear order the 
duality map \eqref{dualitytransform} implies the following identifications 
of operators about the regular boson and critical fermion theories
\begin{equation} \begin{split} \label{id}
({\bar \phi}\phi)=&{- \kappa_F \over 4\pi} \sigma_F,\\
\left( {\bar \phi}\phi \right)^2 =& ({-\kappa_F\over4\pi})^2 \sigma_F^2,\\
\left( {\bar \phi}\phi \right)^3 =& ({-\kappa_F\over4\pi})^3 \sigma_F^3.\\
\end{split}
\end{equation}
Note in particular that the second and third of the identifications 
\eqref{id} simply the square and cube of the first,  
in perfect agreement with the general expectations of large $N$ trace 
factorization. The agreement of \eqref{dualitytransform} with the expectations
of large $N$ factorization constitute a consistency check of these 
transformation formulae.

Let us turn our attention to RG flows beyond linear order.  
Let us first focus on the bosonic theory. The parameter $x_6-1$ multiplies 
a (large $N$) marginal deformation about this theory while the parameters 
$b_4$ and $m_B^2$ multiply relevant operators of dimension two and one 
respectively. Perturbations about the fixed point are characterized by 
two dimensionless numbers $x_6-1$ and $\frac{m_b}{b_4}$ and a dimensionful
number which we can choose to be $b_4$. In the language of the RG flow, 
$x_6-1$ and $\frac{m_b}{b_4}$ may be thought of as `directions' of the flow
lines away from the UV fixed point while $b_4$ represents the renormalization
group scale. In particular the limit $b_4 \to \infty$ captures the deep IR
of the RG flow. In an entirely similar manner $y_6$ is the coefficient of 
a (large $N$) marginal deformation of the critical fermion theory while
$y_4$ and $y_2^2$ are coefficients of relevant operators of dimension two and 
one respectively. Fermionic RG flows may be characterized by two 
dimensionless numbers $y_6$ and $\frac{y_2}{y_4}$ and one scale 
parameter which may be chosen to be $y_4$.

Clearly \eqref{dualitytransform} maps bosonic and fermionic flows 
\begin{equation}\label{bosfermflow}
x_6=1, b_4=x, m_B^2=0,~~~~y_6=0, y_4=x, y_2=0 
\end{equation}
to each other.%
\footnote{ 
In order to argue that this scaling corresponds to an actual RG flow 
one needs to specify a renormalization scheme under the large $N$ limit. 
This should apply to other examples of RG flow given below.
We would like to leave it to future work.   
}
As we have 
discussed above, $x$ is a scale variable along the RG flow. On both sides 
of the duality the limit $x \to \infty$ corresponds to the deep IR of the 
flow. It is not difficult to convince oneself that the 
bosonic Lagrangian \eqref{rst} reduces to that of the critical 
bosonic theory in the limit $x \to \infty$ while the fermionic Lagrangian 
\eqref{csfnonlinear} reduces to that of the regular fermion theory in the 
same limit (see below for details). In other words 
\eqref{bosfermflow}  represent RG flows from the regular boson to the 
critical boson and the critical fermion to the regular fermion theory. 
These flows are mapped to each other under duality for all values of the RG scale 
$x$. 

It is instructive to study two deformations of the critical RG flows  
\eqref{bosfermflow}.
Let us first turn on the marginal parameter on 
both sides, i.e. to study the dual pair of flows  
\begin{equation}\label{bosfermflowd}
x_6-1=a, b_4=x, m_B^2=0,~~~~y_6=\frac{a}{4}, y_4=x, y_2=0 
\end{equation}
for any finite value of the dimensionless number $a$. It is not difficult 
to verify that the deep IR (i.e. limit $x \to \infty$ with $a$ fixed) of these
flows once again reduces to the critical bosonic theory on the bosonic side
and the regular fermion theory on the fermionic side. In other words the 
parameter $a$ is irrelevant in the deep IR of the RG flow. The physical 
reason for this is simple; $x_6-1$ and $y_6$ are coefficients of operators 
whose dimension about the UV fixed point is three, but whose dimension about
the IR fixed point is six. In other words these operators, while 
marginal about the UV fixed point, are highly irrelevant about the 
IR fixed point and so drop out of the IR dynamics of these flows.

The flows \eqref{bosfermflowd} end up at the IR fixed point and so represent
critical flows at all finite values of $a$. Let us now study flows that 
deviate infinitesimally from the critical flow in such a manner that the 
flow ends not in the IR fixed point but in a finite mass deformation about 
this fixed point in the limit $x \to \infty$. In the rest of this 
introduction we find dual pairs of RG flows that have this property.  
We will demonstrate that the end points of these flows are particular 
mass deformations of the critical boson and regular fermion theories that 
have independently been shown to be dual to one another 
(see  \cite{Aharony:2012ns}). We regard this match with the 
previously known duality as a nontrivial consistency check of the duality 
map \eqref{dualitytransform}.

Let us first consider the fermionic theory. Consider the limit%
\footnote{In the critical flow \eqref{bosfermflowd} the dimensionless ratio 
$\frac{y_2^2}{y_4^2}=0$. In the scaling limit \eqref{rf1} the same 
ratio equals  $-2 \frac{m_F^{\rm reg}}{y_4}$. This ratio tends to zero 
in the limit $y_4 \to \infty$, but does so like $\frac{1}{y_4}$. 
As the fermionic mass operator has dimension 2 in the UV but dimension 
1 in the IR, this particular scaling zeroes in on a finite mass deformation 
of the IR theory. Identical comments apply 
to the bosonic RG flow studied below.} 
\beal{ 
y_4 \to \infty, ~~~y_2^2 \to \infty, ~~~y_6, \; \frac{-y_2^2}{2 y_4}=m_F^{\text{reg}}
:{\rm fixed}.
\label{rf1} 
}
Integrating $\sigma_F$ out by using its saddle point equation (a procedure that is justified in the large $N$ limit) we find 
that in this limit $\sigma_F=m_F$; note that in this limit $\sigma_F$ becomes
independent of ${\bar \psi}\psi$.
Under this limit \eqref{csfnonlinear}
reduces to the mass deformed regular fermion theory
\beal{~\label{csfnonlinearnn}
 S_F \to&\int d^3 x \bigg[ i \varepsilon^{\mu\nu\rho} {\kappa_F \over 4 \pi}
\Tr( A_\mu\partial_\nu A_\rho -{2 i\over3}  A_\mu A_\nu A_\rho)
+  \bar{\psi} \gamma_\mu D^{\mu} \psi +m_F^{\text{reg}} {\bar \psi} \psi \bigg]
}
up to a constant term. 

Let us now turn to the bosonic theory. \eqref{dualitytransform} maps the 
limit \eqref{rf1} to 
\begin{equation}
b_4 \to \infty, ~~~m_B^2 \to \infty, ~~~x_6, \;\frac{m_B^2}{2 b_4}=\lambda_B m_B^{\text{cri}}:{\rm fixed}
\label{rbl}
\end{equation}
where we used \cite{Aharony:2012ns,Jain:2013gza}
$$ m_F^{\text{reg}}=-\lambda_B m_B^{\text{cri}}.$$ 
In this limit the term proportional 
to $({\bar \phi} \phi)^3$ in \eqref{rst} can be ignored. The terms proportional 
to $({\bar \phi} \phi)^2$ and $({\bar \phi} \phi)$ can be rewritten as 
$$ \sigma_B ({\bar \phi} \phi) + \alpha_1 \sigma_B + \alpha_2 \sigma_B^2$$
where 
$$
\alpha_1 ={m_B^2\kappa_B\over 8\pi b_4}, \quad \alpha_2 = {-\kappa_B\over 16\pi b_4}.
$$
In the limit \eqref{rbl} the coefficient $\alpha_2$ tends to zero, and 
we obtain the action 
\beal{\label{cbd}
S_B  \to& \int d^3 x  \biggl[i \varepsilon^{\mu\nu\rho} {\kappa_B \over 4 \pi}
\Tr( A_\mu\partial_\nu A_\rho -{2 i\over3}  A_\mu A_\nu A_\rho)
+ D_\mu \bar \phi D^\mu\phi  +\sigma_B({\bar\phi} \phi +N_B\frac{m^{\text{\text{cri}}}_{B}}{4\pi} ) \biggl].
}
In the limit, in other words, the duality \eqref{dualitytransform} reduces 
to the duality between the theories with Lagrangians \eqref{cbd} and 
\eqref{csfnonlinearnn}. But precisely this duality was previously 
conjectured in \cite{Aharony:2012ns,Jain:2013gza}.

\section{Thermal partition functions}

Consider two Chern Simons theories with rank and level 
$(N_B, \kappa_B)$ and $(N_F, \kappa_F)$, which are discrete parameters in the theory, with an identical number of bosonic and fermionic matter fields in the fundamental representation of the U($N_B$) and U($N_F$) gauge group, respectively.
We refer to these 
as the bosonic and fermionic theories respectively. 
Let $p^B_i$ and 
$p^F_i$ denote the continuous parameters in the Lagrangians for these 
two theories. In the context 
of this paper, the bosonic theory is \eqref{rst} and the fermionic 
theory is \eqref{csfnonlinear}.
The parameters $p^B_i$ are ($m_B^2$, $b_4$, 
$x_6$), while the parameters $p^F_i$ are ($y_2^2$, $y_4$, $y_6$).

Consider the partition function of each of these theories on the 
space $S^2 \times S^1$ in the presence of a constant background holonomy 
$U$. Let the circumference of the $S^1$ be given by $\beta$ and the volume of the two sphere by $V_2$. 
$\beta$ is identified with the inverse temperature $T^{-1}$ in the two dimensional theory on $S^2$. 
In the 't Hooft large $N$ limit, under which the two discrete parameters form a new continuous parameter by $\lambda=N/\kappa$, the two partition functions take 
\be 
\exp[-V_2 T^2  v_B(\rho_B(\alpha), \beta, \lambda_B, p^B_i)]
\label{tpfscalar} 
\ee
for the bosonic theory and 
\be 
\exp[-V_2 T^2 v_F(\rho_F(\alpha), \beta, \lambda_F, p^F_i)]
\label{tpffermion} 
\ee
for the fermionc theory. $\rho_B(\alpha)$ is the density of eigenvalues
$e^{i \alpha}$ of the holonomy $U_B$ of the bosonic theory, and 
$\rho_F(\alpha)$ is the density of eigenvalues
$e^{i \alpha}$ of the holonomy $U_F$ of the fermionic theory.%
\footnote{It follows from gauge invariance that the partition functions depend 
only on the set of eigenvalues of the holonomies $U_B$ and $U_F$. Let the 
eigenvalues of, for instance, $U_B$ be given by $e^{i \alpha^B_i }$ for 
$i =1 \ldots N_B$. In the large $N_B$ limit this collection of eigenvalues is 
well characterized by the effectively continuous eigenvalue density 
function 
$$\rho_B(\alpha)= \frac{1}{N_B}\sum_{i=1}^N \delta(\alpha- \alpha^B_i).$$
Identical remarks apply to the fermionic case.}

It was demonstrated in \cite{Jain:2013gza} that the partition function of the 
bosonic
and fermionic theories on an $S^2$ is exactly computed under the large $N$ limit with $\frac{V_2T^2}{N}$ and all other parameters held fixed. For this 
reason we study the thermal free energy of the bosonic and fermionic theories 
on $S^2 \times S^1$ taken to be of order 
$N$. In this limit it was demonstrated in \cite{Jain:2013gza} that the 
$S^2$ partition 
functions of the bosonic and fermionic theories (at all values of $\frac{V_2T^2}{N}$ and theory parameters) agree with each other under 
a proposed duality map of the form 
\begin{equation} \label{fdm}
\kappa_F=-\kappa_B, ~~~N_F=|\kappa_F|-N_B, ~~~\lambda_F=\lambda_B
-{\rm sgn}(\lambda_B), ~~~
p_i^F=p^F_i(p^B_i, \lambda_B)
\end{equation}
if and only if 
\begin{equation}\label{condfordual}
 v_B(\rho_B(\alpha), \beta, \lambda_B, p^B_i)
=  v_F(\rho_F(\alpha), \beta, \lambda_F, p^F_i). 
\end{equation}
The LHS of the equation \eqref{condfordual} is to be evaluated 
under the replacement \eqref{fdm} together with the additional 
replacement 
\beal{
\lambda_F \rho_F(\alpha) = - {\sgn(\lambda_B) \over 2\pi} + \lambda_B \rho_B(\alpha+\pi).
\label{rhotransform}
} 
The rational for the replacement rule \eqref{rhotransform} has its roots in 
the map between Wilson loops in different representations of the gauge group 
under level rank duality and was explained in great detail in 
\cite{Jain:2013py}. In addition as discussed in \cite{Jain:2013py} the leading contribution of the thermal free energy density on $S^2$ under the limit reduces to that on $R^2$. 
We will not pause here to recall the arguments of 
\cite{Jain:2013py}, but simply refer the interested reader to that paper. 

In the rest of this paper 
we test the conjectured duality transformation 
rules \eqref{dualitytransform} as follows. We simply evaluate   
$v_B(\rho_B(\alpha), \beta, \lambda_B, p^B_i)$ and 
$v_F(\rho_F(\alpha), \beta, \lambda_F, p^F_i)$ by direct computation, and 
then explicitly verify that the equation \eqref{condfordual} is 
indeed satisfied once we use the duality map \eqref{dualitytransform}
together with \eqref{rhotransform}

\subsection{Scalars}

The thermal partition function of the deformed regular scalar theory \eqref{rst} has already 
been computed in \cite{Jain:2013gza} (see \cite{Jain:2012qi,Aharony:2012ns,Jain:2013py} for earlier results in 
special cases); in this subsection we present a brief review of these 
results.

The thermal propagator is given by 
\begin{equation} \label{propboson}
\begin{split}
\langle {\phi(p) \bar \phi}(-p')  \rangle &= \frac{(2 \pi)^3 \delta^3(p-p')}
{\wt p^2+ c_B^2 T^2}
\end{split}
\end{equation}
where gauge indices are implicit and 
\be
\wt p_\mu = p_\mu+\delta_{\mu3}\alpha
\label{momentumholonomy}
\ee
with $\alpha$ denoting holonomy. 
$(c_BT)^2$ is the pole mass squared of the scalar fields at the trivial holonomy distribution, which is determined by the gap equation 
\begin{align}
{c_B^2} 
&= (1+3x_6)\lambda_B^2 \cS^2 -4\lambda_B \hat b_4 \cS  +\hat m_B^2
\label{scalargapequation}
\end{align}
where $\hat b_4:= {b_4 \over T}, \hat m_B:= {m_B \over T}$ are dimensionless quantities and we set 
\beal{
\cS :=& \half \int_{-\pi}^\pi d\alpha\rho_B(\alpha)   \( \log(2 \sinh \frac{c_B +i\alpha }{2})+ \log(2 \sinh \frac{c_B - i\alpha }{2}) \) 
}
where we chose $c_B$ as positive in this paper.%
\footnote{ Equivalently saying we denote $|c_B|$ by $c_B$ just for simplicity.  
} 

The thermal free energy normalized in a way of \eqref{tpfscalar} can be obtained by discarding the contribution of the fermions from (2.17) in \cite{Jain:2013gza}:
\beal{
v_{B}[\rho_B] 
=&v_{B,0}+{N_B \over 6\pi} \bigg[- c_B^3 + 2 ({c_B^2} - \hat m_B^2 )\cS  +2 \lambda_B \hat b_4 \cS^2 \nn
&- 3 \int_{-\pi}^\pi d\alpha \rho_B(\alpha) \int_{c_B} ^\infty dy y \( \log (1 - e^{-y-i\alpha }) +\log (1 - e^{-y+i\alpha }) \) \bigg]
\label{scalarTFE}
}
where $v_{B,0}$ is a normalized constant so that $v_B$ goes to zero at zero temperature:
\be
v_{B,0} = -{N_B \over 6\pi} (-\hat c_{B,0}\hat m_B^2+{\lambda_B \over 2}\hat b_4\hat c_{B,0}^2)
\label{vb0}
\ee
where $c_{B,0}$ is the pole mass at zero temperature, $c_{B,0}=\lim_{T\to0}(c_B T)$, and $\hat c_{B,0}$ is normalized by temperature so as to be dimensionless: $\hat c_{B,0}:={c_{B,0}\over T}$.%
\footnote{ Note that $T\cS \to {c_{B,0}\over 2}$ under $T\to0$.}  
Note that the gap equation \eqref{scalargapequation} satisfies the  saddle point equation extremizing the thermal free energy in terms of $c_B$. 

\subsection{Fermions}

We now turn to new computation of this note. 
We follow a method developed in \cite{Giombi:2011kc,Jain:2012qi,Yokoyama:2012fa}
to compute the thermal free energy in $U(N_F)_{\kappa_F}$ Chern Simons theory with fermions in the fundamental representation. Although the computation is 
a straight-forward application of the methods outlined in the references above, we present some details of the formalism and our computations for the 
convenience of readers. 

\subsubsection{Zero temperature} 

In order to illustrate our techniques in a simple setting, we first review 
the computation of the pole mass of fermions at zero temperature. 
In the next subsection we will quickly be able to generalize the formulae 
of this subsection to the physically interesting finite temperature case.

We start with the Lagrangian \eqref{csfnonlinear} defined on $ R^3$. 
We first take the (Euclidean) light-cone gauge $A_-:={A_1\pm iA_2 \over \sqrt2}=0$,%
\footnote{This gauge may be justified by analytic continuation with $i\epsilon$-prescription \cite{Giombi:2011kc}. See \cite{Moshe:2014bja} for computation in a different gauge. }
so that the gauge self-interaction disappears.
As a result we can integrating out the gauge field except its zero mode or holonomy. 
The result is given by
\beal{
S_F
=& \int  \frac{d^3 p}{(2\pi)^3} [\bar{\psi}(-p) i\gamma^\mu \wt p_\mu \psi(p)  +\sigma_F\bar\psi(-p) \psi(p)]  - {V N_F y_2^2\over4\pi\lambda_F}\sigma_F -   {V N_F y_4\over4\pi\lambda_F} \sigma_F^2 +  {V N_F y_6\over4\pi\lambda_F} \sigma_F^3  \nn
&+N_F \int \frac{d^3P}{(2 \pi)^3} \frac{d^3q_1}{(2 \pi)^3} \frac{d^3q_2}{(2 \pi)^3}~
\frac{8 \pi i\lambda_F}{ (q_1-q_2)_{-}} \xi_-(P,q_1) \xi_I(-P,q_2),
\label{sf1}
}
where $V = (2\pi)^3 \delta^3(P=0)$, $\wt p_\mu$ is given by \eqref{momentumholonomy} and we set 
\beal{
\xi_I(P,q) := \frac{1}{2N_F}\bar{\psi}(\frac{P}{2}-q) \psi(\frac{P}{2}+q), \quad
\xi_-(P,q) := \frac{1}{2N_F}\bar{\psi}(\frac{P}{2}-q) \gamma_-\psi(\frac{P}{2}+q).
\label{xi}
}
Here we already assumed the auxiliary field $\sigma_F$ to be constant in order to study vacuum structure of this theory. 
Notice that the 2nd line in \eqref{sf1} is quartic fermionic interaction generated by integrating out the gauge field.

Then we introduce bilocal auxiliary fields denoted by $\alpha_I, \alpha_-, \Sigma^I, \Sigma^-$ and consider terms 
\beal{
S_{\text{aux}} 
=& - N_F\int \frac{ d^3P}{(2 \pi)^3}\frac{d^3 q}{(2 \pi)^3} \( 2\Sigma^I(-P, q) (\alpha_I(P,q) - \xi_I(P,q)) + 2\Sigma^-(-P, q) (\alpha_-(P,q) - \xi_-(P,q)) \)\nn
& - N_F \int\frac{ d^3P}{(2 \pi)^3}\frac{ d^3q_1}{(2 \pi)^3} \frac{d^3q_2}{(2 \pi)^3}~\frac{8 \pi i\lambda_F}{(q_1-q_2)_{-}} \xi_-(P,q_1) \xi_I(-P,q_2) \nn
&+ N_F \int\frac{ d^3P}{(2 \pi)^3}\frac{ d^3q_1}{(2 \pi)^3} \frac{d^3q_2}{(2 \pi)^3}~
\frac{8 \pi i\lambda_F}{(q_1-q_2)_{-}} \alpha_-(P,q_1) \alpha_I(-P,q_2)
}
which gives no dynamical effect, since evaluating this by integrating out the auxiliary fields $\Sigma^I, \Sigma^-$ gives trivial result. 
We add this term into the action given in \eqref{sf1} to cancel the quartic fermionic interaction.
\beal{
&S_F+S_{\text{aux}} \nn
=& \int  \frac{d^3 p}{(2\pi)^3} \bar{\psi}(-p) \( i\gamma^\mu \wt p_\mu +\sigma_F \)  \psi(p)  - {V N_F y_2^2\over4\pi\lambda_F}\sigma_F -   {V N_F y_4\over4\pi\lambda_F} \sigma_F^2 +  {V N_F y_6\over4\pi\lambda_F} \sigma_F^3  \nn
&- N_F\int \frac{ d^3P}{(2 \pi)^3}\frac{d^3 q}{(2 \pi)^3} \( 2\Sigma^I(-P, q) (\alpha_I(P,q) - \xi_I(P,q)) + 2\Sigma^-(-P, q) (\alpha_-(P,q) - \xi_-(P,q)) \)\nn
&+ N_F \int\frac{ d^3P}{(2 \pi)^3}\frac{ d^3q_1}{(2 \pi)^3} \frac{d^3q_2}{(2 \pi)^3}~
\frac{8 \pi i\lambda_F}{(q_1-q_2)_{-}} \alpha_-(P,q_1) \alpha_I(-P,q_2).
}
Now this is quadratic in terms of fermionic fields, one can integrate them out by gaussian integration. And we impose translation invariance for the gauge-singlet bilocal fields, since we are interested in vacuum configuration.
\beal{
&\Sigma^I(P,q) = (2\pi)^3 \delta^3(P) \Sigma^I(q) , \quad
\Sigma^-(P,q) = (2\pi)^3 \delta^3(P) \Sigma^-(q), \\
&\alpha_I(P,q) = (2\pi)^3 \delta^3(P) \alpha_I(q) , \quad
\alpha_-(P,q) = (2\pi)^3 \delta^3(P) \alpha_-(q) .
}
After this manipulation we find  
\bes{
S_F+S_{\text{aux}}
=& N_FV \biggl[ \int  \frac{d^3 p}{(2\pi)^3} {1\over N_F} \Tr_{G,S}[ \log \( i\gamma^\mu \wt p_\mu +\sigma_FI+\Sigma(p) \)^{-1}] - { y_2^2\over4\pi\lambda_F}\sigma_F -   {y_4\over4\pi\lambda_F} \sigma_F^2 +  {y_6\over4\pi\lambda_F} \sigma_F^3  \\
&+\int \frac{d^3q_1}{(2 \pi)^3} \frac{d^3q_2}{(2 \pi)^3}~
\frac{8 \pi i\lambda_F}{ (q_1-q_2)_{-}} \alpha_-(q_1) \alpha_I(q_2)  - \int \frac{d^3 q}{(2 \pi)^3} \(2 \Sigma^I( q) \alpha_I(q) +   2 \Sigma^-( q) \alpha_-(q) \) \biggl]
\label{exactSeff}
}
where we set 
$
\Sigma(p) =\Sigma^{-}(p)\gamma_{-}+\Sigma^{I}(p)I
$
and $\Tr_{G,S}$ represents the trace for fundamental gauge indices as well as for spinor ones. 
Notice that \eqref{exactSeff} reaches the canonical form of the exact effective action in the leading of large $N$, which is written only in terms of singlet fields with $N$ factored out (except holonomy contribution). 

Saddle point equations for $\Sigma, \alpha, \sigma_F$ are 
\bes{ 
\alpha(q)=& - {1\over N_F} \Tr_{G}[ {1\over  i\gamma^\mu \wt q_\mu +\wt\Sigma(q) }  ], \\
{\Sigma}^{-}(p)=&{4 \pi i \lambda_F}\int \frac{d^3q}{(2\pi)^3}\frac{1}{(p-q)_{-} } \alpha_I(q),\\
{\Sigma}^{I}(p)=&-{4\pi i \lambda_F}\int \frac{d^3q}{(2\pi)^3} \frac{1}{(p-q)_{-}} \alpha_-(q),\\
0=&\int  \frac{d^3 q}{(2\pi)^3} {1\over N_F}\Tr_{G,S} [{-1 \over i\gamma^\mu \wt q_\mu +\wt\Sigma(q) } ]
-{y_2^2\over4\pi\lambda_F}- {y_4\over2\pi\lambda_F}\sigma_F +{3y_6\over4\pi\lambda_F} \sigma_F^2,
\label{gapeqs}
}
where $\Tr_{G}$ means taking trace only in fundamental gauge indices, and we set 
$
\alpha(p) := \alpha_{-}(p)\gamma^{-}+\alpha_{I}(p)I
$
and 
$
\wt \Sigma(p) := \Sigma(p) +\sigma_F I.
$

Note that the auxiliary field $\wt\Sigma$ becomes the exact self energy in the 't Hooft large $N$ limit: 
\begin{equation} \label{propfermion}
\begin{split}
\aver{\psi(p)\bar\psi(-p') } &= { (2 \pi)^3 \delta^3(-p'+p)\over i\gamma^\mu \wt p_\mu + \wt \Sigma(p)}
\end{split}
\end{equation}
where gauge and spinor indices are abbreviated. 

\subsubsection{Finite temperature}

Introduction of temperature can be done in a standard way by compactifying one direction with a fixed circumference denoted by $\beta$, which is interpreted as the inverse temperature.
Fermions in thermal canonical ensemble obey the anti-periodic boundary condition for this circle. 
Then the momentum of a fermion of the $S^1$ direction is discretized as a half integer:
$p_{3} = (n+\half)\beta^{-1}$, where we use the 3rd direction as the compactified one.
Thus the computation and the results we have done at zero temperature can be used as those at finite temperature by replacing the momentum in the 3rd direction by discretized one. For example, the integration measure of momentum is replaced in a way that 
\be
\int {d^3 p \over (2\pi)^3} \to \int {\cD^3 p \over (2\pi)^3}:={1\over \beta} \sum_{p_{3} = (n+\half)\beta^{-1} \atop n\in\bf Z} \int {d^2 p \over (2\pi)^2}.
\label{replace} 
\ee

By doing this manipulation, the saddle point equation for $\Sigma$ at finite temperature is obtained from \eqref{gapeqs} as 
\bes{ 
{\Sigma}^{-}(p)=&{4 \pi i \lambda_F}\int \frac{\cD^3q}{(2\pi)^3}\frac{1}{(p-q)_{-} } {-1\over 2N_F} \Tr_{G,S}[ {1\over  i\gamma^\mu \wt q_\mu +\wt\Sigma(q) }  ],\\
{\Sigma}^{I}(p)=&-{4\pi i \lambda_F}\int \frac{\cD^3q}{(2\pi)^3} \frac{1}{(p-q)_{-}}{-1\over 2N_F} \Tr_{G,S}[\gamma_- {1\over  i\gamma^\mu \wt q_\mu +\wt\Sigma(q) }  ].
\label{gapeqsigma}
}
These saddle point equations can be solved in the same manner as done in \cite{Giombi:2011kc,Jain:2012qi,Yokoyama:2012fa}. 
The fact that the light-cone gauge preserve rotational symmetry on the two-plane enables one to set ansatz such that
\be
\wt\Sigma_{F,I}(p) = f(\hat p) p_s, \quad 
\Sigma_{F,+}(p) =i g(\hat p) p_+,
\label{ansatzsigmaf}
\ee
where $f(\hat p), g(\hat p)$ are undetermined functions of $\hat p={p_s \over T}$, which is dimensionless. Dependence of other parameters is implicit.
Then one can show from \eqref{gapeqsigma} that there exists a dimensionless constant $c_F$
such that 
\be
f(\hat p)^2 + g(\hat p) = {c_F^2 \over \hat p^2}. 
\label{fgc}
\ee
Plugging this back into \eqref{gapeqsigma} gives the integral equation for $f(\hat p)$ only, which can be solved as follows.
\bes{
f(\hat p) =&{\lambda_F \over \hat p} \int_{-\pi}^\pi d\alpha\rho_F(\alpha)   \( \log(2 \cosh \frac{\sqrt{\hat p^2 +c_F^2} +i\alpha}{2})+ \log(2 \cosh \frac{\sqrt{\hat p^2 +c_F^2} - i\alpha}{2}) \)+\hat \sigma_F.
}
In the intermediate step we took the large $N$ limit, in which holonomy eigenvalues distribute densely in the region $[-\pi,\pi]$ so that the summation over a gauge index becomes integral over this holonomy distribution denoted by $\rho(\alpha)$:
\be
{1\over N} \sum_{i=1}^N F(\alpha_i) \overset{N\to\infty}{\to} \int_{-\pi}^\pi d\alpha \rho(\alpha) F(\alpha)
\ee
where $F(\alpha_i)$ is an arbitrary function of eigenvalues of holonomy $\alpha_i$.
$g(\hat p)$ is determined from \eqref{fgc} so that $g(\hat p) = {c_F^2 \over \hat p^2} - f(\hat p)^2$. Since \eqref{gapeqsigma} indicates that $g(\hat p)$ is free from IR divergence, it has to be satisfied that 
$\lim_{\hat p\to0}({c_F^2} - \hat p^2f(\hat p)^2)=0 $, 
which gives determining equation for $c_F$: 
\be
c_F^2 =({2\lambda_F} \cC +\hat\sigma_F)^2,
\label{cf}
\ee
where we set $\hat \sigma_F := {\sigma_F \over T}$ and 
\be
\cC := \half \int_{-\pi}^\pi d\alpha\rho_F(\alpha)   \( \log(2 \cosh \frac{c_F +i\alpha }{2})+ \log(2 \cosh \frac{c_F - i\alpha }{2}) \). 
\ee
The equation \eqref{cf} is invariant under the flip of signature of $c_F$, so we can choose $c_F$ as positive without losing generality.
We will call \eqref{cf} the gap equation of fermion because $c_F$ represents the pole mass (normalized by temperature) of fermions when the holonomy distribution is trivial, which can be seen from 
\eqref{propfermion} by using $\det (i\gamma^\mu p_\mu +\wt\Sigma) = p_\mu^2+c_F^2T^2$. 

On the other hand, $\sigma_F$ is determined from the last equation in \eqref{gapeqs}, which can be computed as 
\beal{
-3 \hat\sigma_F^2 y_6-4 \hat\sigma_F \lambda_F\cC+2 \hat\sigma_F \hat y_4-4 \lambda_F\cC^2+\hat y_2^2=0
\label{sigmaeq}
}
where $\hat y_4= {y_4 \over T}, \hat y_2 = {y_2 \over T}$.%
\footnote{ 
The solutions are given by
\beal{
\hat\sigma_F = { \hat y_4 - 2\lambda_F \cC \pm \sqrt{( \hat y_4 -  2\lambda_F\cC)^2 -  {3 y_6} ((2\lambda_F\cC)^2-{\hat y_2^2})} \over 3 y_6}.
\label{sigmasol}
}
}

The thermal free energy without normalization is obtained by carrying out the replacement \eqref{replace} for the exact effective action given by \eqref{exactSeff}: 
\bes{
F=& N_FV \biggl[ \int  \frac{\cD^3 p}{(2\pi)^3} {1\over N_F} \Tr_{G,S}[ \log \( i\gamma^\mu \wt p_\mu +\wt\Sigma(p) \)^{-1}] - { y_2^2\over4\pi\lambda_F}\sigma_F -   {y_4\over4\pi\lambda_F} \sigma_F^2 +  {y_6\over4\pi\lambda_F} \sigma_F^3  \nn
&+\int \frac{\cD^3q_1}{(2 \pi)^3} \frac{\cD^3q_2}{(2 \pi)^3}~
\frac{8 \pi i\lambda_F}{ (q_1-q_2)_{-}} \alpha_-(q_1) \alpha_I(q_2)  - \int \frac{\cD^3 q}{(2 \pi)^3} \(2 \Sigma^I( q) \alpha_I(q) +   2 \Sigma^-( q) \alpha_-(q) \) \biggl].
}
Following the computation done in \cite{Giombi:2011kc,Jain:2012qi,Yokoyama:2012fa,Jain:2013gza} we can compute this as follows. 
\beal{
F=& {N_F V_2 T^2 \over 6 \pi}  \biggl[c_F^3 - 2{\lambda_F^2 } \cC^3 - {3  \over 2} ({c_F^2} - \hat\sigma_F^2) \cC - {3\hat y_2^2\over 2\lambda_F}\hat \sigma_F -   {3\hat y_4\over2\lambda_F}\hat\sigma_F^2 +  {3y_6\over2\lambda_F}\hat\sigma_F^3 \nn
& - 3 \int_{-\pi}^\pi d\alpha \rho_F(\alpha) \int_{c_F}^\infty dy y \( \log (1 + e^{- y-i\alpha} ) +\log (1 + e^{- y+i\alpha } ) \) \biggr]
}
where we used $V=V_2\beta$. 
Therefore the free energy with the normalization of \eqref{tpffermion} is given by
\beal{
v_F[\rho_F] 
=&v_{F,0}+ {N_F\over 6 \pi}  \biggl[c_F^3 - 2{\lambda_F^2 } \cC^3 - {3  \over 2} ({c_F^2} - \hat\sigma_F^2) \cC  - {3\hat y_2^2\over 2\lambda_F}\hat \sigma_F -   {3\hat y_4\over2\lambda_F}\hat\sigma_F^2 +  {3y_6\over2\lambda_F}\hat\sigma_F^3\nn
& - 3 \int_{-\pi}^\pi d\alpha \rho_F(\alpha) \int_{c_F}^\infty dy y \( \log (1 + e^{- y-i\alpha} ) +\log (1 + e^{- y+i\alpha } ) \) \biggr],
\label{fermionTFE} 
}
where $v_{F,0}$ is a normalized constant, which we determine by requiring $v_F \to 0$ under $T\to0$.
We will give an explicit form later, \eqref{vf0}. 

Let us determine an {\it off-shell} form of the free energy density in the sense that extremizing it in terms of $c_F, \sigma_F$ gives the gap equations \eqref{cf}, \eqref{sigmaeq}.  
For this purpose we eliminate $\cC$ by using the gap equation of $c_F$, \eqref{cf}, or square-root of it, which is given by 
\beal{
c_F = \sgn[2\lambda_F \cC + \hat \sigma_F] (2\lambda_F \cC + \hat \sigma_F).
\label{cflinear}
}
Note that we already chose $c_F$ as positive. 
Then the saddle point equation of $\sigma_F$ becomes 
\beal{
-3 \hat\sigma_F^2 y_6+\hat\sigma_F^2+2 \hat\sigma_F \hat y_4-c_F^2+\hat y_2^2=0
\label{sigmaeq2}
}
and the thermal free energy density is written as 
\beal{
v_F[\rho_F] 
=&v_{F,0}+ {N_F\over 6 \pi}  \biggl[c_F^3(1 -{\sgn[2\lambda_F \cC + \hat \sigma_F]\over\lambda_F}) + {1\over2\lambda_F}({3\hat\sigma_Fc_F^2-\hat\sigma_F^3} - {3\hat y_2^2}\hat \sigma_F -   {3\hat y_4 }\hat\sigma_F^2 +  {3y_6}\hat\sigma_F^3)\nn
& - 3 \int_{-\pi}^\pi d\alpha \rho_F(\alpha) \int_{c_F}^\infty dy y \( \log (1 + e^{- y-i\alpha} ) +\log (1 + e^{- y+i\alpha } ) \) \biggr]. 
\label{fermionTFEoffshell} 
}
It is not difficult to see that extremizing this thermal free energy density in terms of $c_F, \sigma_F$ yields the same saddle point equations as \eqref{cflinear}, \eqref{sigmaeq2}, respectively.  

The gap equation obtained above simplifies in the zero temperature limit. 
In this limit the gap equation becomes%
\footnote{ We used $T\cC \to {c_{F,0}\over 2}$ under $T\to0$.} 
\bes{
&c_{F,0}^2 =(\lambda_F c_{F,0}+ \sigma_{F,0} )^2 
}
where $\sigma_{F,0}$ is determined by 
\beal{ 
&-3\sigma_{F,0}^2 y_6+ \sigma_{F,0}^2+2  \sigma_{F,0}   y_4-c_{F,0}^2+  y_2^2=0.
\label{gapeqzero}
}

\section{Duality} 

In this section we demonstrate that the gap equations and the thermal 
partition functions of the bosonic and fermionic theories map to each 
other under the parameter map  \eqref{dualitytransform} as well 
together with the holonomy distributions given by \eqref{rhotransform}.

\subsection{Duality of the gap equation}

We first demonstrate that the fermionic gap equation maps to that of the bosonic one under the duality transformations. 
For this purpose, we remove the $\sigma_F$ from the equations \eqref{cf} and \eqref{sigmaeq}.  
By using \eqref{cflinear}, \eqref{sigmaeq} becomes 
\beal{
&-3 (\sgn[2\lambda_F \cC + \hat \sigma_F] c_F -2 \lambda_F \cC)^2 y_6-4 (\sgn[2\lambda_F \cC + \hat \sigma_F] c_F -2 \lambda_F \cC) \lambda_F\cC \nn
&+2 \hat\sigma_F \hat y_4-4 \lambda_F\cC^2+\hat y_2^2=0
}
which can be simplified as 
\be
-3 c_F^2 y_6+2 c_F \sgn[2\lambda_F \cC + \hat \sigma_F] (\hat y_4-2 \lambda_F\cC+6 \lambda_F\cC y_6)-4 \lambda_F\cC (\hat y_4-\lambda_F\cC+3 \lambda_F\cC y_6)+\hat y_2^2=0.
\ee
Now let us perform the known duality map of holonomy density function given by \eqref{rhotransform}, which can be used as a form such that%
\footnote{The inverse relation is 
$
\lambda_B \cS = -{ \sgn(\lambda_F) \over 2 } c_F +\lambda_F \cC.
$
}
\beal{
\lambda_F \cC = -{ \sgn(\lambda_B) \over 2 } c_B +\lambda_B \cS.
\label{cCtocS} 
}
Then we obtain 
\beal{
&c_F^2 (2 \sgn(\lambda_B) \sgn[2\lambda_F \cC + \hat \sigma_F](3 y_6-1)+6 y_6-1)+4 (\lambda_B \cS) (\hat y_4-(\lambda_B \cS)+3 (\lambda_B \cS) y_6)-\hat y_2^2 \nn
&-2 c_F (\sgn(\lambda_B)+\sgn[2\lambda_F \cC + \hat \sigma_F]) (\hat y_4-2 (\lambda_B \cS)+6 (\lambda_B \cS) y_6) =0. 
\label{dualcf2}
}
Under a situation where 
\be
\sgn[2\lambda_F \cC + \hat \sigma_F] =\sgn(\lambda_F)
\label{condition}
\ee
\eqref{dualcf2} reduces to 
\beal{ 
c_F^2+4 {\hat y_4} (\lambda_B \cS)-4 (\lambda_B \cS)^2-\hat y_2^2+12 (\lambda_B \cS)^2 y_6=0
}
where we also used $\sgn(\lambda_B)=-\sgn(\lambda_F)$. One can easily see that this transformed gap equation of the fermions is precisely identical to that of scalar theory \eqref{scalargapequation} by the duality relation \eqref{dualitytransform} with $c_B = c_F$. 

\subsection{Duality of the free energy}

We can also show that the thermal free energy of fermions precisely maps that of bosons under the duality relations. 
To this end we rewrite the thermal free energy density given by \eqref{fermionTFE} by using \eqref{cf} and \eqref{sigmaeq} so that 
\beal{
v_F[\rho_F] 
=&v_{F,0}+ {N_F\over 6 \pi}  \biggl[c_F^3 -2 {c_F^2}  \cC +{1\over2\lambda_F}(-\hat y_4\hat\sigma_F^2 -2\hat y_2^2\hat\sigma_F) \nn
& - 3 \int_{-\pi}^\pi d\alpha \rho_F(\alpha) \int_{c_F}^\infty dy y \( \log (1 + e^{- y-i\alpha} ) +\log (1 + e^{- y+i\alpha } ) \) \biggr], 
\label{fermionTFE2} 
}
where $v_{F,0}$ is given by 
\beal{
v_{F,0} =&- {N_F\over 6 \pi\lambda_F}(-\half\hat y_4\hat\sigma_{F,0}^2 - \hat y_2^2\hat\sigma_{F,0})
\label{vf0}
}
with $\sigma_{F,0}$ determined from \eqref{gapeqzero} and $\hat \sigma_{F,0}:={\sigma_{F,0}\over T}$. 
Under a situation in \eqref{condition}, we can rewrite $\sigma_F, \sigma_{F,0}$ as 
\beal{ 
\hat\sigma_F = -2\lambda_B \cS, \quad 
\sigma_{F,0} =-\lambda_B c_{B,0}
}
by using \eqref{cflinear} and \eqref{cCtocS}. 
Thus the thermal free energy of the deformed critical fermionic theory divided by $\kappa_F$ is rewritten as 
\beal{
{v_F[\rho_F] \over \kappa_F} 
=&{v_{F,0}\over \kappa_F}+ {1\over 6 \pi}  \biggl[\lambda_F c_F^3 -2 {c_F^2} \lambda_F \cC +{1\over2}(-\hat y_4\hat\sigma_F^2 -2\hat y_2^2\hat\sigma_F) \nn
& - 3 \int_{-\pi}^\pi d\alpha \lambda_F\rho_F(\alpha) \int_{c_F}^\infty dy y \( \log (1 + e^{- y-i\alpha} ) +\log (1 + e^{- y+i\alpha } ) \) \biggr]\nn
=&{v_{F,0}\over \kappa_F}+ {1\over 6 \pi}  \biggl[(\lambda_B-\sgn(\lambda_B)) c_B^3 -2 {c_B^2} (-{\sgn(\lambda_B)\over2}c_B+\lambda_B \cS) +{1\over2}(-\hat b_4 (-2\lambda_B \cS)^2 -2\hat m_B^2(-2\lambda_B \cS)) \nn
& - 3 \int_{-\pi}^\pi d\alpha (-{\sgn(\lambda_B) \over \pi}+\lambda_B\rho_B(\alpha+\pi) ) \int_{c_B}^\infty dy y \( \log (1 + e^{- y-i\alpha} ) +\log (1 + e^{- y+i\alpha } ) \) \biggr]\nn
=&{v_{F,0}\over \kappa_F}+ {1\over 6 \pi}  \biggl[\lambda_B c_B^3 -2 {c_B^2}\lambda_B \cS -2\hat b_4 (\lambda_B \cS)^2 +2\hat m_B^2(\lambda_B \cS)) \nn
& - 3 \int_{-\pi}^\pi d\alpha \lambda_B\rho_B(\alpha)  \int_{c_B}^\infty dy y \( \log (1 -e^{- y-i\alpha} ) +\log (1 - e^{- y+i\alpha } ) \) \biggr], 
\label{dualfermionTFE} 
}
with 
\beal{ 
{v_{F,0}\over \kappa_F} =& - {1\over 6 \pi}(-\half\hat y_4\hat\sigma_{F,0}^2 - \hat y_2^2\hat\sigma_{F,0})
=- {1\over 6 \pi}(-\half\hat b_4(-\lambda_B c_{B,0})^2 - \hat m_B^2(-\lambda_B c_{B,0})) \nn
=&- {\lambda_B\over 6 \pi}(-\half\lambda_B \hat b_4c_{B,0}^2 +\hat m_B^2c_{B,0}). 
\label{dualvf0}
}
Comparing \eqref{scalarTFE}, \eqref{vb0} we find 
\be
{v_F[\rho_F] \over \kappa_F}  = -{v_B[\rho_B] \over \kappa_B}. 
\ee
That is, the thermal free energies of two theories are identical by using $\kappa_F = -\kappa_B$.
This completes our demonstration of the proposed duality. 

\subsection{Comments on the duality} 
\label{comments}

In this subsection we give a comment on Hubbard-Stratonovich-like approach to obtain the duality proposed in this paper by adding auxiliary fields from a known duality pair \cite{Gur-Ari:2015pca}.%
\footnote{ 
We would like to thank the referee for noticing to us that the method of the reference \cite{Gur-Ari:2015pca}, which appeared on the arXiv at the same date as this paper, can be applied to the current case, irrespective of the issue to deal with contact terms in the method. 
}
To see this
begin with a simplest duality pair of the regular boson and critical fermion theories, whose parameters are chosen as $m_B^2=b_4=0,x_6=1$ and $y_2^2=y_4=y_6=0$, respectively. 
It is known that these are dual to each other with the operator mapping $\bar\phi \phi = {- \kappa_F \over 4\pi} \sigma_F$, which can be seen from comparison of correlation functions.
Then, instead of turning on the parameters illustrated in Introduction in this paper, 
one may add to those actions the following term constituted by auxiliary fields $D_1, D_2$ such that 
\beal{
\Delta S = \int d^3 x [\cO D_1 - D_1 D_2 + \beta_1 D_2 + \beta_2 D_2^2 + \beta_3 D_2^3]
}
where $\cO = \bar\phi \phi$ for the bosonic theory and $\cO ={- \kappa_F \over 4\pi} \sigma_F$ for the fermionic one, and $\beta_1, \beta_2, \beta_3$ are parameters. 
It is natural to expect that 
this addition preseves bosonization duality with the same operator mapping. 
Integrating out the auxiliary fields gives 
\beal{
\Delta S = \int d^3 x [\beta_1 \cO + \beta_2 \cO^2 + \beta_3 \cO^3]
}
The total bosonic fermionic action leads to \eqref{rst} with the identification $\beta_1 = m_B^2, \beta_2 = {4\pi b_4 \over \kappa_B}, \beta_3 = { (2\pi)^2 (-1 + x_6) \over \kappa_B^2}$, and the fermionic one becomes \eqref{csfnonlinear} by $\beta_1 = y_2^2, \beta_2 =-{4\pi y_4 \over \kappa_F}, \beta_3 = -{ (4\pi)^2 y_6 \over \kappa_F^2}$. 
This is consistent with the parameter mapping \eqref{dualitytransform} obtained
by comparing the thermal free energies of both sides.
This method provides a simple check of the result.

\section{Discussion}

In this paper we have conjectured a simple and explicit duality between 
two RG flows. The first of these is the flow from the large $N$  free scalar theory to the critical scalar theory. The second is the flow from 
the large $N$ critical fermion theory to the free fermionic one. The $U(N)$ 
symmetry group in both the RG flows studied above are gauged and gauge 
dynamics is governed, in each case, by a Chern Simons term. The duality 
between these RG flows holds provided the level and rank on the two 
sides of this duality are related by the usual rules of level rank duality. 
The actual duality map between Lagrangian parameters, 
\eqref{dualitytransform}, turned out to be linear in contrast with the 
complicated nonlinear map of \cite{Jain:2013gza}. It would be interesting 
to understand the reason for this simplicity. 

If we accept that the two RG flows above are dual to each other we really have only one flow. 
This flow may picturesquely be described as the flow from a free boson 
theory to a free fermion theory - with the provision, of course that 
each of these theories is Chern Simons gauged with different levels and ranks. 

It would be useful to find additional evidence for our conjectured duality. 
One way might be to compare the S matrices of the two theories imitating 
the analysis of \cite{Jain:2014nza} and \cite{Inbasekar:2015tsa}.
Especially from S matrices one can extract information of possible bound states of the theory by analyzing poles of S matrices \cite{Dandekar:2014era}, which are to be mapped to each other between the dual theories under duality transformation \cite{Inbasekar:2015tsa}. 
Together with S. Jain and M. Mandlik we have made a preliminary attempt 
at the relevant computations. We find that the S matrices of the bosonic 
and fermionic theories are indeed identical upto a subtlety. The 
computations of the S matrix in the fermionic theory turns out to include 
a contribution from one divergent integral.%
\footnote{
This is in contrast with the study of \cite{Jain:2014nza} in which all S matrix 
integrals were finite. The new divergence appears to have its roots in the 
fact that a term of the form $\psi^4$ is not power counting renormalizable.}
The fermionic and bosonic S matrices match perfectly provided we 
assign this integral a value different from the one obtained from dimensional
regularization. We do not yet understand the rational for the ad hoc cut off
scheme that leads to the duality invariant value. We hope to return to this 
question in the future.

To end this paper,  let us note that the RG flows 
\eqref{bosfermflow} have a very simple dual bulk interpretation.
Recall that the dual regular boson and critical fermion theories 
have both been conjectured to be dual to the $AdS_4$ solutions of 
Vasiliev's equations. Adopting the `bosonic representation', 
the Vasiliev system in question is the Type A theory deformed by 
the interaction phase $e^{ i \frac{\pi \lambda}{2}}$ 
\cite{Giombi:2011kc, Chang:2012kt}. The boundary conditions on the 
scalar field in the Vasiliev multiplet are set so that the dual scalar 
operator has unit dimension (i.e so that Vasiliev scalar field dies off 
near the boundary like $z$; the coefficient of the order $z^2$ fall off 
vanishes). 

In the bosonic representation, the RG flow \eqref{bosfermflow} is simply 
a deformation of the regular scalar theory by the double trace operator
$ ( \bar \phi \phi)^2$. According to the general rules of the AdS/CFT 
correspondence, however, in the large $N$ limit the addition of a double 
trace operator to the boundary theory does not modify the bulk solution; 
it simply modifies the boundary conditions of the corresponding bulk scalar 
\cite{Witten:2001ua, Berkooz:2002ug}. The deformation by $b_4=x$ changes 
the boundary condition on the bulk scalar field so that it dies off 
like $ A z + B z^2$ at infinity, where the 
ratio $\frac{B}{A}$ is a function of $x=b_4$ that vanishes when $b_4 \to 0$, 
but diverges as $b_4 \to \infty$ (see \cite{Chang:2012kt} for a 
careful and detailed derivation of the precise form of the boundary 
condition). In the deep UV of the RG flow the boundary condition is simply 
that appropriate to the regular scalar theory we started with. In the deep 
IR, on the other hand, the boundary condition for Vasiliev bulk scalar field 
is that it must die off at infinity like $z^2$ (the coefficient of the 
$z$ fall off vanishes). But this is precisely the boundary condition 
of the Vasiliev system dual to the critical bosonic or regular fermionic 
scalar theory. It follows that the modified boundary conditions described 
above do indeed have the properties expected of RG flow \eqref{bosfermflow}, 
namely that they interpolate between the dual to the regular scalar theory and 
the critical scalar theory as $b_4$ varies from zero to infinity. 
See \cite{Klebanov:2002ja, Klebanov:1999tb} for closely related discussions.

Above we have presented explicit formulae for the free energy 
of the bosonic (and fermionic) theories at every value of $b_4$ and temperature.
It should presumably be possible to reproduce our explicit formula for the 
free energy from the thermodynamics of black brane solutions in Vasiliev 
theory subject to the appropriate boundary conditions. It would certainly be 
very interesting to perform this check. It is, however, not yet clear 
whether the required solutions of Vasiliev's equations are known or 
how their thermodynamical properties can be extracted once they are
determined. For this reason we leave the intensely interesting 
comparison between the thermodynamical formulae in this paper and those 
of black branes in Vasiliev theory to future work.


\section*{Acknowledgments}
We would like to thank S. Jain and M. Mandlik for collaboration at the early 
stage of this project. We would also like to thank T. Takimi, V. Umesh, 
S. Wadia and E. Witten for helpful discussions.  
S.M. would like to thank the 
University of Caltech for hospitality while this work was initiated. The work 
of S.M. was 
supported in part by a joint UGC-ISF (Indo Israel) grant. 
The work of S.Y. was supported in part by the Israeli Science Foundation
under grant 504/13.
S.M. would like to 
acknowledge his debt to the people of India for their generous and steady support for research in basic sciences.

\bibliographystyle{utphys}
\bibliography{csdeformv2}

\end{document}